\documentclass[showpacs,preprintnumbers,amsmath,amssymb,epsfig,widetext,twocolumn]{revtex4}

\usepackage{graphicx}% Include figure files
\usepackage{dcolumn}% Align table columns on decimal point
\usepackage{bm}% bold math
\usepackage{epsfig}

\def\fun#1#2{\lower3.6pt\vbox{\baselineskip0pt\lineskip.9pt
  \ialign{$\mathsurround=0pt#1\hfil##\hfil$\crcr#2\crcr\sim\crcr}}}

\input epsf

\def\be{\begin{equation}}
\def\ee{\end{equation}}
\def\ba{\begin{eqnarray}}
\def\ea{\end{eqnarray}}

\def\nn{\nonumber}

\begin{document}

\preprint{}

\title{Consistency test of general relativity from large scale
structure of the Universe}

\author{Yong-Seon Song \footnote{Yong-seon.Song@port.ac.uk}
and Kazuya Koyama \footnote{Kazuya.Koyama@port.ac.uk}}

\affiliation{Institute of Cosmology $\&$ Gravitation,
University of Portsmouth, Portsmouth, PO1 2EG, UK }

\date{\today}

\begin{abstract}
We construct a consistency test of General Relativity (GR) on cosmological
scales. This test enables us to distinguish between the two alternatives
to explain the late-time accelerated expansion of the universe, that is,
dark energy models based on GR and modified gravity
models without dark energy. We derive the consistency relation in
GR which is written only in terms of observables - the Hubble parameter,
the density perturbations, the peculiar velocities and the lensing potential.
The breakdown of this consistency relation implies that the Newton constant
which governs large-scale structure is different from that in the background
cosmology, which is a typical feature in modified gravity models.
We propose a method to perform this test by reconstructing the
weak lensing spectrum from measured density perturbations and peculiar
velocities. This reconstruction relies on Poisson's equation in GR
to convert the density perturbations to the lensing potential. Hence any
inconsistency between the reconstructed lensing spectrum and the measured
lensing spectrum indicates the failure of GR on cosmological scales.
The difficulties in performing this test using actual observations
are discussed.
\end{abstract}

%\pacs{draft}

%\keywords{CMB-inflation}

\maketitle
\section{Introduction}
 In 1998, astronomers discovered that the expansion of the Universe was accelerating, not
slowing down as expected~\cite{Riess:1998cb, perlmutter98}.
This late-time acceleration of the Universe is surely the most
challenging problem in cosmology. Within the framework of General Relativity (GR), the
acceleration originates from dark energy. The simplest option is the cosmological
constant, first introduced by Einstein. However, in order to explain the current
acceleration of the Universe, the required value of the cosmological constant must be
incredibly small. Particle physics predicts the existence of vacuum energy, but it is
typically many orders of magnitude larger than the observed values of the
cosmological constant (for a review see Ref.~\cite{Copeland:2006wr}).

Alternatively, there could be no dark energy, but a large-distance modification to GR
may account for the late-time acceleration of the universe~\cite{dvali00,carroll03} (for
a review see \cite{Nojiri:2006ri, Koyama:2007rx}).
In order to realize the late-time acceleration, it is necessary to modify GR at
cosmological scales. It is really challenging to construct a consistent theory of modified
gravity (MG) and progress in theoretical physics is required in order to have
fully consistent models. Even if successful theoretical models for MG are constructed,
we need to distinguish them from dark energy models in GR
via observations. This is indeed possible
if one can combine various data sets which measure not only the expansion history
of the Universe but also
the structure of the Universe~\cite{Lue:2003ky}. However,
it has been argued that by fine-tuning the properties of dark energy, it is always
possible to mimic any MG model and hence it is impossible to distinguish
between the two possibilities~\cite{Kunz:2006ca,Bertschinger:2008zb}.

In this paper, we argue that it is indeed {\it possible} to distinguish
between MG models and dark energy models, and we propose a
consistency test of GR which enables us to distinguish between these two possibilities
(see also Ref.~\cite{Jain:2007yk}).
From large-scale structure in the Universe, we can
measure three quantities. One is the density perturbation which can be measured from
the galaxy distribution and cluster abundance. The second is the peculiar velocity
which can be measured from the redshift distortions of the galaxy power spectrum and
the internal dynamics of clusters and galaxies.
Finally, we can measure the lensing potential from weak lensing
of galaxies and clusters. We construct a consistency equation in GR which can be written
only in terms of these observables. In this way, it is possible to avoid using
specific MG models or parametrising MG models in order to test GR on cosmological scales.
Then we discuss how we can realize this test using actual observations in the future.
The most difficult part of the consistency test is that all the quantities should be
measured at the same time and at the same location. However, weak lensing only measures
the integrated effects of the lensing potential at different redshifts and there is no simple
way to reconstruct the lensing potential at a given redshift. In this paper,
we propose a novel way to overcome this problem by reconstructing the lensing
power spectrum from measured matter perturbations at given redshifts.

Before going into the construction of the consistency test, we first summarize the
assumptions that we make in this paper. First of all, we assume
the cosmological principle and describe our background universe as an isotropic
and homogeneous Friedmann-Robertson-Walker universe. Thus we do not consider the possibility of
an inhomogeneous universe which has been extensively studied in the literature as
a possible explanation for the late-time accelerated expansion of the Universe
(see \cite{Buchert:2007ik} for a review).
The cosmological principle can be tested from
observations \cite{Goodman:1995dt, Caldwell:2007yu, Clarkson:2007pz, Uzan:2008qp}
and we assume that we can use this to exclude the inhomogeneous
universe. Next, we assume that the energy-momentum tensor for dark matter
and baryons is conserved.
In MG models, we try to
explain the late-time accelerated expansion of the Universe without introducing
any exotic fluids, so this is a reasonable assumption. In dark energy models,
dark matter can be coupled to dark energy (see for example \cite{Amendola:1999er}).
However, coupling between baryons and dark energy is strongly constrained by
local experiments. Thus only the conservation of
the energy-momentum tensor for dark matter can be broken. This creates a difference
between peculiar velocities of dark matter and baryons, which can be tested from observations.
This test will be investigated in a forthcoming paper \cite{SKM} and we assume that we
can use this to exclude the possibility of the breakdown of energy-momentum
conservation for dark matter. The final assumption is that MG models are metric
theories and there exists a Newtonian limit of the theory.
This is necessary to reproduce the very tight constraints on deviations from GR
at solar system scales.
With this assumption, on subhorizon scales, only the time-time component and the anisotropic
part of the space-space component of the modified Einstein equations for perturbations are
important. From the Bianchi identity and the conservation of the energy-momentum tensor, other components of the modified Einstein equations should be trivially satisfied.

The structure of this paper is as follows. In section II, we summarize the
basic equations for background cosmology and perturbations in
GR and MG. In section III, the consistency test of GR is constructed,
written only in terms of observables. We provide a set of consistency
equations depending on the assumptions for the properties of dark energy in GR.
In section IV, we propose a method to perform the test by reconstructing the
weak lensing power spectrum from measured matter perturbations.
Section V is devoted to conclusions.

\section{Background and perturbations}
In GR, the Friedman equation determines the expansion rate of the Universe
\begin{equation}
H^2 = \frac{8 \pi G}{3} \rho^{GR}_t, \quad \rho^{GR}_t = \rho_m + \rho_{de},
\label{FRGR}
\end{equation}
where $\rho_m$ is the matter energy density which include both dark matter, and
baryon and the dark energy density $\rho_{de}$
accounts for the late-time acceleration of the Universe.

On the other hand, in MG models, the expansion
is determined by a modified Friedman equation $H^2= F(8 \pi G \rho_m)$ where
$F$ is a function determined by the underlying modified theory of gravity.
We can always rewrite this modified Friedman equation as
\begin{equation}
H^2 =  \frac{8 \pi \bar{G}_{eff}(t)}{3} \rho^{MG}_t,
\quad \rho^{MG}_t = \rho_m.
\label{MGFR}
\end{equation}

For a given MG model, it is always possible to find a dark energy
model which gives exactly the same expansion history, i.e. the same $H$.
This gives the relation between total energy densities in two models as
$\rho^{GR}_t/\rho^{MG}_t = \bar{G}_{eff}(t)/G$.

We use the Newtonian gauge to describe the metric perturbations
\begin{equation}
ds^2 = - (1+2 \Psi) dt^2+ (1+2 \Phi) a^2 \delta_{ij}dx^i dx^j.
\end{equation}
The density fluctuation is defined by
$\delta = (\rho -\bar{\rho})/\bar{\rho}$ for each component of matter
where the quantities with bar denotes the background quantity.
The divergence of the peculiar velocity is
$\theta = \nabla_j T^j_0 /(\bar{\rho} + \bar{P})$.
In GR, the perturbed Einstein equations at sub-horizon scales give
\begin{eqnarray}
k^2 \Phi &=& 4 \pi G a^2 \rho_t^{GR}
\delta_t^{GR}, \\
k^2 (\Phi + \Psi) &=& - 12 \pi G a^2 (1+w_t) \rho_t^{GR}
\sigma_t^{GR},
\end{eqnarray}
where $\sigma$ is an anisotropic stress. The quantity with
the subscript t is the total contribution from matter
perturbations
\begin{eqnarray}
\rho_t \delta_t &=& \rho_m \delta_m + \rho_{de} \delta_{de}, \\
 (\rho_t + P_t) \theta_t &=& (\rho_{de} + P_{de}) \theta_{de}
+\rho_m \theta_m, \\
(\rho_t + p_t) \sigma_t &=& (\rho_{de} + \rho_{de}) \sigma_{de}.
\end{eqnarray}
As explained in the introduction, we assume that there is no interaction
between dark energy and dark matter/baryons. Then conservation of the
energy-momentum tensor gives
\begin{equation}
\dot{\delta}_m = -\frac{\theta_m}{a}, \quad
\dot{\theta}_m = -H \theta_m + \frac{k^2 \Psi}{a},
\label{conmatter}
\end{equation}
and
\begin{eqnarray}
\dot{\delta}_{de} &=& -(1+w_{de}) \frac{\theta_{de}}{a}
- 3 H \frac{\delta P_{de}}{\rho_{de}} + 3 H
w_{de} \delta_{de}, \;\;\;\;\;\;\; \\
\dot{\theta}_{de} &=& - H (1-3w_{de}) \theta_{de}
-\frac{\dot{w}_{de}}{1 + w_{de}} \theta_{de} \nonumber\\
&& + \frac{k^2}{a}
\left(\frac{1}{\rho_{de}} \frac{1}{1+w_{de}} \delta P_{de}
-\sigma_{de} + \Psi \right),
\end{eqnarray}
where $\delta P_{de}$ is the pressure perturbation
of the dark energy fluid.

Assuming a Newtonian limit in the MG model,
the modified Einstein equations yield the following equations \cite{Jain:2007yk}
\begin{eqnarray}
k^2 \Phi &=& 4 \pi G_{eff}(a, {\bf k})a^2  \rho^{MG}_t
\delta_t^{MG},
\label{MGPO}
\\
\Phi &=& - \eta(a ,{\bf k}) \Psi,
\end{eqnarray}
where $G_{eff}(a, {\bf k})$ and $\eta(a, {\bf k})$
should be determined by the modified theory of gravity.
In the MG model, we have only dark matter ($\rho_t = \rho_m$,
$\delta_t= \delta_m$) and the conservation equations are given by
Eq.~(\ref{conmatter}). All other components of modified Einstein
equations should be satisfied trivially thanks to the Bianchi
identity.

\section{Consistency test of GR}
First, let us consider simple dark energy models like quintessence,
where we can safely ignore clustering of
dark energy, at least under horizon scales. In this case,
the metric perturbations are given by
\begin{equation}
k^2 \Phi^{GR} = 4 \pi G a^2 \rho^{GR}_m \delta_m^{GR}, \quad
\Psi^{GR} = -\Phi^{GR},
\end{equation}
in the GR dark energy model.
In reality there is no direct measurement of the curvature
perturbation $\Phi$. The weak gravitational lensing provides the
combination $\Phi_- \equiv (\Phi - \Psi)/2$ which
determines the geodesics of photons. Then
we can define a consistency parameter as
\begin{equation}
\label{testI}
\alpha^{(1)}(a, {\bf k}) = \frac{k^2 \Phi_-}{4 \pi G a^2 \rho_m \delta_m},
\end{equation}
where $G$ is the Newton constant measured by local experiments.
If GR holds, $\alpha^{(1)}=1$ (see also Ref.~\cite{Uzan:2000mz} for
an early attempt to test GR from large-scale structure and gravitational
lensing). We emphasize that all quantities in the
consistency relation are observables and we do not need to assume any theory.
It is easy to see that unless $(1+\eta^{-1}) G_{eff} /2G=1$,
we have $\alpha^{(1)} \neq 1$ in MG. We should emphasize that this
test can allow a special class of MG with
$(1+\eta^{-1}) G_{eff} /2G=1$ to pass the GR consistency test $\alpha^{(1)}=1$.
However, if $G_{eff}/G \neq 1$, then $\eta \neq 1$, which means that
the GR relation $\Phi =-\Psi$ is modified. In order to check this, we need one more
equation which relates the Newtonian potential $\Psi$ to matter-energy
perturbations. Since energy-momentum is conserved,
the continuity equation for matter is not altered and
Eq.~(\ref{conmatter}) gives $k^2 \Psi = d (a \theta_m)/dt$.
Then we define a consistency parameter for $\eta \neq 1$ as
\begin{equation}
\tilde{\alpha}^{(1)}(a, {\bf k})=-\frac{\Phi}{\Psi}
= -\frac{8 \pi G a^2 \rho_m \delta_m}{3\,a^2 H^2V_m}\,,
\label{conpi}
\end{equation}
%\begin{equation}
%\alpha^{\pi}(a, {\bf k})=\Psi+ \Phi = \frac{3}{2}\left(\frac{a^2 H^2}{k^2}\right)
%V_m + \frac{4 \pi G a^2 \rho_m \delta_m}{k^2},
%\label{conpi}
%\end{equation}
where $V_m= 2 (d\theta_m/dt + H \theta_m)/(3 aH^2)$.
$V_m$ can be constructed by direct measurements of the peculiar velocity
and acceleration of matter. We should emphasize that only
energy-momentum conservation for dark matter is used. Thus even if
there is a peculiar velocity of dark energy $\theta_{de}$, the reconstruction
of the Newtonian potential is not affected. If $\alpha^{(1)}=1$ and
$\tilde{\alpha}^{(1)}=1$, we can conclude that there is no modification of GR
and dark energy has no clustering.

However, even if one finds $\alpha^{(1)} \neq 1$ or $\tilde{\alpha}^{(1)} \neq 1$,
this does not give definite evidence for the breakdown of GR. In general,
dark energy can have non-trivial clustering.
In fact, the conservation equations for dark energy perturbations
are not closed and the pressure perturbation $\delta P_{de}$ and the
anisotropic stress $\sigma_{de}$ should be determined by the microphysics
of the dark energy fluid. Given that we do not know the origin
of dark energy, we can treat these two functions as arbitrary
functions. Then it is always possible to mimic $\Phi$ and
$\Psi$ in a MG model by tuning  $\delta P_{de}$ and
$\sigma_{de}$ in a dark energy model in GR, because $\delta \rho_{de}$
mimics $G_{eff}$ and $\sigma_{de}$ mimics $\eta$ \cite{Kunz:2006ca}.
This fact can be clearly seen by rewriting Poisson's equation
in GR with clustering DE as
\begin{eqnarray}
k^2 \Phi &=& 4 \pi G a^2 (\rho_m \delta_m + \rho_{de} \delta_{de})\nonumber\\
&=& 4 \pi G a^2 \left(1+ \frac{\rho_{de} \delta_{de}}
{\rho_m \delta_m} \right) \rho_m \delta_m.
\end{eqnarray}
Then, if we only observe dark matter fluctuations $\delta_m$, we
cannot distinguish between clustering DE and modified gravity
as the clustering DE mimics the modification of Newton's constant
\begin{equation}
G_{eff}(a, {\bf k}) = G \left(1+ \frac{\rho_{de} \delta_{de}}{\rho_{m} \delta_{m}} \right).
\end{equation}
Thus if we find $\alpha^{(1)} \neq 1$ and $\tilde{\alpha}^{(1)} \neq 1$, there are
still two possibilities. One is MG and the other is dark energy model
with non-trivial clustering.

However, this argument is clearly based on the assumption that we only
observe the clustering of dark matter~\cite{Jain:2007yk}.
Once the dark energy clusters like dark matter, both perturbations
participate in forming halos and thus galaxies. Thus the galaxy distribution
should trace the total matter perturbations, which are the
sum of the dark matter and dark energy perturbations. In the same way,
the cluster abundance is sensitive to the total matter perturbations
\cite{Jain:2007yk}. When we measure the total matter perturbations
from the galaxy distribution, bias between the galaxy density
fluctuations and the total matter perturbations becomes the main
issue. In this paper, we assume that bias can be measured
independently by higher order statistics or other methods.
We will come back to this issue in the conclusions.

Assuming we can measure the total density perturbations
from observations, it becomes possible to improve the consistency
test and we can distinguish clustering dark energy and modified gravity.
Including the dark energy perturbations, Poisson's equation is now written as
\begin{equation}
k^2 \Phi^{GR} = 4 \pi G a^2 \rho_t^{GR} \delta_t^{GR},
\end{equation}
in the GR dark energy model, and
\begin{equation}
k^2 \Phi^{MG} = 4 \pi G_{eff}(a, {\bf k}) a^2 \rho_t^{MG} \delta_t^{MG},
\end{equation}
in MG models. Let us consider two models that give the same $\alpha^{(1)}(a, {\bf k}) \neq 1$.
In these two models, $\Phi^{GR}=\Phi^{MG}$. Then the
relation between total density perturbations in the two models is
\begin{equation}
\frac{\delta_t^{GR}}{\delta_t^{MG}} = \frac{G_{eff} \rho_t^{MG}}{G \rho_t^{GR}}
= \frac{G_{eff}(a, {\bf k})}{\bar{G}_{eff}(a)}.
\label{deltat}
\end{equation}
Here $\bar{G}_{eff}(a)$ is the effective Newton constant determined
by the background Friedman equation (\ref{MGFR}) and $G_{eff}(a, {\bf k})$ is
the effective Newton constant determined by the Poisson equation (\ref{MGPO}).
In general these two effective Newton constants are
different, as is seen in explicit examples like the DGP \cite{dvali00}
and $f(R)$ gravity models \cite{carroll03}. We expect that this is a very general
feature of MG.

Then we can define the consistency parameter as
\begin{equation}
\alpha^{(2)}(a, {\bf k}) = \frac{k^2 \Phi}{4 \pi G a^2 \rho_t \delta_t}.
\end{equation}
In GR, $\alpha^{(2)}=1$. The total energy density $\rho_t$ is reconstructed from $H$
using the Friedman equation in GR, Eq.~(\ref{FRGR}).  In this case, we cannot assume the GR relation $\Phi=-\Psi$ and
we should combine weak lensing that measures $\Phi_-$ and
peculiar velocity that determines $\Psi$. Then
the consistency parameter is given by
\ba\label{eq:GRcon}
\alpha^{(2)}(a, {\bf k}) =
\frac{4 k^2}{3 a^2}
\left( \frac{\Phi_- - \frac{3}{4}a^2H^2V_m}{H^2 \delta_t} \right)\,.
\ea
Again all the quantities in this equation can be measured from observations
and we do not need to assume any theory to perform the test. From
Eq.~(\ref{deltat}) and using the fact that in clustering DE and MG,
$\Phi_-$, $V_m$ and $H$ are the same, we find
\begin{equation}
\alpha^{(2)}(a, {\bf k})=\frac{G_{eff}(a, {\bf k})}{\bar{G}_{eff}(a)},
\end{equation}
in MG models. This test (\ref{eq:GRcon}) is essentially the same as the consistency
relation proposed in \cite{Jain:2007yk}. Here, we have shown that this consistency
relation is a test of the difference between the Newton constants for the
background cosmology and for sub-horizon perturbations.

\section{Consistency test from projected power spectrum}
All observables in the consistency test should be measured
at the same time and at the same location. However, in reality,
there is no simple method to compare the statistical observables
$\Phi_-$, $\delta_t$ and $V_m$ in the consistency equation
Eq.~(\ref{eq:GRcon}). This is because weak lensing is an
integrated effect of the lensing potential determined by $\Phi_-$
along the line of sight and it is impossible to reconstruct the lensing
potential at a given redshift. In this paper, we propose a novel
way to accomplish the test using the projected power spectrum
(see \cite{Lewis:2006fu} for a review of the projected
power spectrum).

The projected angular power spectra $C^{XX'}_{\ell}$ of any pair of
perturbations $X$ and $X'$ are given by
\ba\label{eq:cldd}
C_{\ell}^{XX'}=\frac{2\pi^2}{\ell^3}\int dDD W^X(D) W^{X'}(D)\frac{9}{25}
\Delta_{\zeta\zeta},\;\;
\ea
where $D$ is the comoving distance and
we used the Limber approximation which is valid for large $\ell$.
$\Delta_{\zeta\zeta}(a_0,k)$ is the rms amplitude of curvature fluctuations
on comoving hypersurfaces at some time given by $a=a_0$ during the matter-dominated
era. The window function $W^X(D)$ is determined by the property of the quantity
$X$ as is shown below.

The deflection angle ${\bf d}$ due to gravitational lensing is defined by
the gradient field of the lensing potential,
${\bf d}={\bf \nabla} \phi$, where
\ba
\phi=-2\int dD \frac{D_s-D}{DD_s}\Phi_-,
\ea
and $D_s$ denotes the
comoving distance to the source galaxies distributed on the thin
redshift shell labeled by $s$.
Then the window function for $\phi$ is
\ba
W^{s\,\phi}(D)=-2 {\cal G}_{\Phi_-}(a,k)\frac{(D_s-D)}{D_sD},
\ea
where the growth function $ {\cal G}_{\Phi_-}(a,k)$ is given by
${\cal G}_{\Phi_-}(a,k)=\Phi_-(a,k)/\Phi_-(a_0,k)$.
The angular power spectrum of the deflection angle is given by
$C_{\ell}^{s\,dd}=\ell(\ell+1)C_{\ell}^{s\,\phi\phi}$,
and $C_{\ell}^{s\,dd}$ is related to the convergence power spectrum
$C_{\ell}^{s\,\kappa\kappa}$ as $C_{\ell}^{s\,\kappa\kappa}
=\ell(\ell+ 1)C_{\ell}^{s\,dd}/4$.

The density perturbations are measured at a given redshift labeled by $i$
at the comoving distance $D_i$ from the observer.
In the approximation of the quasi-static evolution of perturbations,
the projected angular power spectrum can be written in discretized form
at the given redshift bin $i$ as
\be
C_{\ell}^{i \; XX'} = \frac{2 \pi^3}{l^3}
\Delta D_i D_i W^X(D_i) W^{X'}(D_i)\frac{9}{25}\Delta_{\zeta\zeta}.
\ee
The galaxy density fluctuation $\delta_g$ is a biased tracer of the total density
perturbation $\delta_t$, which includes the possible presence of dark energy
clustering: $\delta_g =b \delta_t$. The window function for the $\delta_g$ component
is given by
\ba
W^{i\,\delta_g}(D_i)=\frac{2}{3} {\cal G}_{\delta_t}(a_i,k)
\frac{dz_i}{dD}n_i b_i
\frac{l^2}{\Omega_mH_0^2D_i^2}\,,
\ea
where $n_i$ is the number density of galaxies at
$z=z_i$, $n_g(z_i)$ and $b_i$ is $b(z_i)$.
The growth function ${\cal G}_{\delta_t}$ is given by
\begin{equation}
{\cal G}_{\delta_m}(a_i,k)=\frac{a_i \Phi(a_i,k)}{\Phi(a_0,k)}
\end{equation}
if there is no clustering of dark energy ($\delta_t =\delta_m$)
and
\begin{equation}
{\cal G}_{\delta_t}(a_i,k)=\frac{\Omega_mH_0^2\Phi(a_i,k)}{a_i^2H^2(a_i) \Phi(a_0,k)},
\end{equation}
if there is clustering of dark energy. Both growth functions are normalized
to $a_0$ during the matter-dominated era.

The redshift-space power spectrum of galaxies is anisotropic
due to the peculiar velocities of galaxies. We use an approximation
that peculiar velocities of galaxies trace that of matter $\theta_m$
despite the existence of dark energy peculiar velocity
$\theta_{de}$ \cite{Jain:2007yk}. This allows
a statistical measurement of the peculiar velocities
$\theta_m$ (see for example \cite{Hamilton:1997zq}).
It is possible to extract the power spectrum of $\theta_m$
independently of galaxy bias \cite{Will08}. The derivative of
$\theta_m$ appearing in $V_m$ can be derived from a direct measurment of
the acceleration field~\cite{Kaiser:1987qv}, or estimated from measured
peculiar velocities at different time slicings~\cite{David08}.
Here we assume that it is built from one of these methods.
Then the window function for $V_m$ in each bin $i$ is
\ba
W^{i\,V_m}(D_i)=\frac{2}{3} {\cal G}_{V_m}(a_i,k)\frac{dz_i}{dD_i}n_i
\frac{\ell^2}{\Omega_mH_0^2D_i^2}\,, \;\;
\ea
where the growth function ${\cal G}_{V_m}(a_i,k)$ is given by
\begin{equation}
{\cal G}_{V_m}(a_i,k)=\frac{\Omega_mH_0^2 \Psi(a_i,k)}{a_i^2H^2(a_i) \Psi(a_0,k)}.
\end{equation}

The galaxy power spectra $C_{\ell}^{i\,\delta_g\delta_g}$
and $C_{\ell}^{i\,V_mV_m}$ can be compared directly at each $\ell$
because these quantities are measured at the same time for a given
measured comoving distance $D_i$, and at the same scale
$k=\ell/D_i$ in the Limber approximation.
However, we can observe only the integrated effect of the lensing
potential and it is impossible to measure $\Phi_-$ at a given time
and location. Thus we propose a new way to perform the consistency test.
We statistically reconstruct the lensing power spectrum
$\tilde{C}_{\ell}^{s\,dd}$ by replacing the lensing potential $\Phi_-$ with
the measured density perturbations $\delta_t$ and peculiar velocities $V_m$
at given redshifts.
In the reconstruction, we use Poisson's equation in GR to relate $\delta_t$
to the 3D curvature perturbations $\Phi$. Thus if there is inconsistency between
$C_{\ell}^{s\,dd}$ and $\tilde{C}_{\ell}^{s\;dd}$, it indicates a modificaiton of
the GR Poisson equation, and hence the breakdown of GR in structure formation.

We illustrate how this idea works in the following.
The continuous contributions from the lensing potential in Eq.~(\ref{eq:cldd})
can be discretized
\ba\label{eq:disccldd}
C_{\ell}^{s\,dd}
=\frac{2\pi^2}{\ell}\sum_{i=1}^{n}\Delta D_i D_i
\frac{4(D_s-D_i)^2}{D_s^2D_i^2}\Delta_{\Phi_-\Phi_-}(a_i,k). \;\;\;
\ea
The first test is to assume that there are no dark energy perturbations.
Then $\Delta_{\Phi_-\Phi_-}(a_i,k)$ can be replaced by the measured
matter perturbations from the galaxy distribution
\begin{equation}
\label{eq:phi_phi_}
\Delta_{\Phi_-\Phi_-}^{\;i}=\frac{9}{8\pi^2 \ell }\frac{D_i^3}{\Delta D_i}
\left(\frac{dz}{dD}n_ib_i\right)^{-2}
\frac{\Omega_m^2H_0^4}{a_i^2}C_{\ell}^{i\,\delta_g\delta_g}.
\end{equation}
Substituting this into Eq.~(\ref{eq:disccldd}), we derive the
{\it reconstructed} power spectrum
\begin{eqnarray}\label{recon_m}
\tilde{C}_{\ell}^{\; s\,dd}(\alpha^{(1)})&=&\frac{9}{\ell^2}
\sum_{i=1}^{n}\frac{D_i^2(D_s-D_i)^2}{D_s^2}
\left(\frac{dz}{dD}n_ib_i\right)^{-2} \nonumber\\
&& \times
\frac{\Omega_m^2H_0^4}{a_i^2}C_{\ell}^{i\,\delta_g\delta_g}.
\end{eqnarray}
In the upper panel of Fig.~1, we demonstrate how accurately
this reconstruction can be done, using LCDM with
the cosmological parameters $\Omega_m=0.25$, $\Omega_b=0.05$,
$H_0=72$ km/sec/Mpc, and a flat model with no DE clustering.
The primordial perturbations are characterized by $\Delta_{\zeta\zeta}(a_0,k)=
\delta_{\zeta}^2(a_0,k_n)(k/k_n)^{(n-1)}T^2(k)$,
where $\delta_{\zeta}(a_0,k_n)=4.52\times 10^{-5}$, $n=0.95$,
$k_n=0.05 {\rm Mpc}^{-1}$, and $T(k)$ denotes the transfer function.
The bin number of source galaxy distributions is
running from 1 to 6 between $z=0.25$ and $z=2.75$, with the spacing
$\Delta z_s=0.5$. For definiteness, we assume that the galaxy sets come
from a net galaxy distribution of $n_g(z)\propto z^2 e^{-(z/1.5)^2}$.
%and assume the precise redshift measurements.
The biasing $b(z_i)$ is fixed to 1. We show the accuracy of this reconstruction
with diverse binning spacing of $\Delta z_i = 0.02$ and $0.1$.

\begin{figure}[t]
  \begin{center}
  \epsfysize=4.5truein
  \epsfxsize=3.truein
    \epsffile{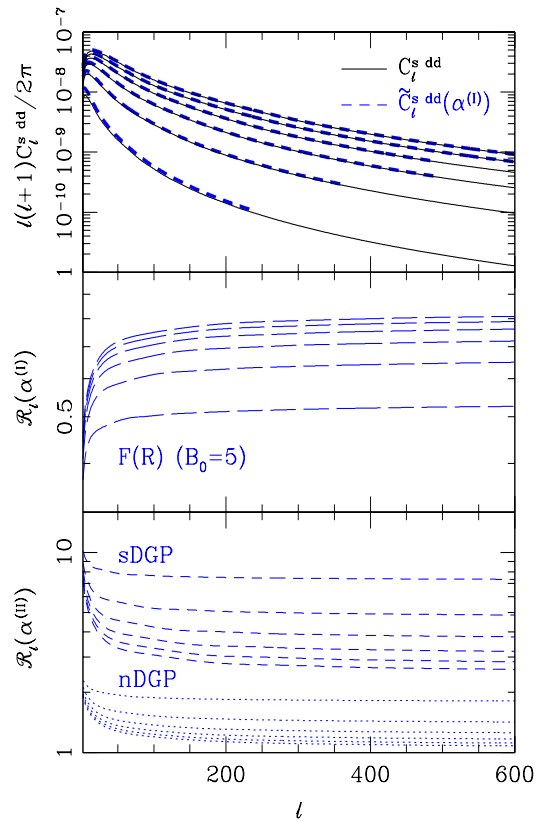}
    \caption{\footnotesize
{\it Upper panel}: The precision test between $C_{\ell}^{s\,dd}$ (solid curves)
and $\tilde{C}_{\ell}^{s\,dd}$
with a redshift spacing $\Delta z_i = 0.1$ (dash curves)
for 6 different $z_s$
running from 0.25 (bottom) to 2.75 (top).
We restricted the reconstructed power spectra to scales where the non-liner 
effects give less than $5\%$ changes on the power spectra. 
{\it Middle panel}: The first consistency test for
GR model (solid curves) and MG model -f(R) gravity~\cite{Song:2006ej} (long dash curves).
{\it Bottom panel}: The second consistency test for
GR model (solid curves), sDGP model~\cite{lue04,koyama05} (dash curves),
and nDGP model~\cite{Song:2007wd,Cardoso:2007xc}(dotted curves).}
    \label{fig:alpha}
  \end{center}
\end{figure}

We define an estimator corresponding to the test Eq.~(\ref{testI}) as
\ba\label{eq:caseI}
{\cal R}_{\ell}(\alpha^{(1)})=
\frac{\tilde{C}_{\ell}^{s\,dd}(\alpha^{(1)})}{C_{\ell}^{s\,dd}}\,.
\ea
We can also define an estimator which provides a test of anisotropic stress
or $\eta =1$ as
\ba\label{eq:anisotest}
{\cal R}^i_{\ell}(\tilde{\alpha}^{(1)})=
\frac{\Omega_m^2H_0^4C_{\ell}^{i\,\delta_g\delta_g}}{a_i^6
H(a_i)^4C_{\ell}^{i\,V_mV_m}}\,.
\ea
If the consistency conditions $\alpha^{(1)}=1$ and $\tilde{\alpha}^{(1)}=1$
are broken, these estimators deviate from $1$. Then there
are two possibilities left: MG or dark energy with clustering and/or
anisotropic stress. In this case, we should reconstruct the lensing potential
using $\delta_g$ and $V_m$. Then $\Delta^i_{\Phi_-\Phi_-}$ in the
discretized approximation is given by
\ba\label{eq:phi_phi_II}
\Delta_{\Phi_-\Phi_-}^{i}&=&\frac{9}{32\pi^2 l}\frac{D_i^3}{\Delta D_i}
\left(\frac{dz}{dD}n_i \right)^{-2} a_i^4H^4(a_i) \\
&\times&\left[b_i^{-2} C_{\ell}^{i\,\delta_g\delta_g}
+b_i^{-1} 2C_{\ell}^{i\,\delta_gV_m}+C_{\ell}^{i\,V_mV_m}\right]\,.\nn
\ea
The reconstructed deflection power spectra $\tilde{C}_{\ell}^{s\,dd}(\alpha^{(2)})$
is given by putting this into Eq.~(\ref{eq:disccldd}).
Then the estimator for the final consistency test Eq.~(\ref{eq:GRcon}) is given by
\ba\label{eq:caseII}
{\cal R}_{\ell}(\alpha^{(2)})
= \frac{\tilde{C}_{\ell}^{s\,dd}(\alpha^{(2)})}{C_{\ell}^{s\,dd}}.
\ea
In the middle/bottom panel of Fig.~1,
we show ${\cal R}_{\ell}(\alpha^{(2)})$
in a $f(R)$ gravity model and ${\cal R}_{\ell}(\alpha^{(2)})$ in the DGP model
as examples for MG models. Clearly, the deviation from
the GR consistency relation is larger than the errors coming from
the reconstruction. Of course, in order
to apply this test to real observations, there are numbers of problems we
should solve, such as the precise measurements of redshifts, the
reconstruction of $V_m$ and the measurements of bias for $\delta_t$.
We should also refine our formulation
including redshift distortions.
However, it is encouraging that it is in principle possible to distinguish
between two major possibilities for the late-time acceleration.
It would be crucial to calculate the signal to noise ratio in future
observations and optimize the detection of the deviation from the consistency
condition.

\section{Systematic errors in consistency test}
The main challenge in realizing the consistency test is to understand bias
and the distribution of galaxies. One possibility to measure bias is
to use higher-order statistics proposed by Ref.~\cite{Matarrese:1997sk}
and applied to the 2dF \cite{Verde:2001sf}. Assuming
a local bias $\delta_g =b_1 \delta_m + b_2 \delta_m^2/2$ and
gaussian initial conditions, the measured bias in the 2df is
$b=1.04 \pm 0.11$. For the SDSS LRGs, Ref.~\cite{Kulkarni:2007qu}
reported that the measured bias is $b=1.87 \pm 0.07$ assuming a linear
biasing. Ref.~\cite{Sefusatti:2007ih} gives a forecast for the measurements of bias
from the bispectrum. Assuming primordial gaussian fluctuations,
a $100$ Gpc$^3$ survey could measure $b_1$ with a few percent accuracy. We
should note that there are additional systematic errors coming from the
assumption on the local biasing. The other method is to use the redshift
distortions to extract the velocity divergent power
spectrum $P_{\theta_m \theta_m}$ which we explain below \cite{Song:2008xd}.
In the first test, we assume that dark energy has no perturbations and the growth
of structure is determined solely by the background cosmology. Then we can reconstruct
the power spectrum of density perturbations $\delta_m$ using
Eq.~(\ref{conmatter}), from which we can measure bias by
comparing it with the galaxy power spectrum $P_{\delta_g \delta_g}$.
Ref.~\cite{Song:2008xd} reported that a  full sky survey would measure
bias at a sub-percent level at $z<2$. This includes the systematic errors in the
measurement of the velocity divergent power spectrum
discussed below. In general, there are additional systematic
errors because the assumptions made to measure bias in
the above arguments should be modified depending on the
nature of gravity. Thus it is very important
to understand the effect of MG on bias \cite{Hui:2007zh}.
One possibility is to extend the halo model approach \cite{Mo:1995cs}
to MG models. Here the modified Newton constant affects the critical
density $\delta_c$ at which the spherical over-density
collapses. The study of spherically symmetric collapse in modified gravity
has just begun \cite{Schaefer:2007nf} and it is necessary to understand the
relation between $G_{eff}(a, {\bf k})$ and $\delta_c$ which eventually
determines the halo bias. The uncertainty in galaxy redshift distributions
also introduces inconsistency between the reconstructed lensing power spectrum
and measured lensing spectrum. This is studied in detail in~\cite{Song:2008xd}.
It is found that the uncertainty in redshift measurements give an
extra bias to the measurements of density perturbations.

In Ref.~\cite{Song:2008xd}, the errors in the reconstructed power spectrum coming from
the uncertainty in the measurement of bias and redshift distributions are estimated
as
\begin{equation}
\Delta \tilde{C}_{\ell}^{s \; dd} = \left\{
\sum^n_{i=1} \left[\frac{1}{b_i^2} F_{\ell}^i \left(2 \frac{\Delta b_i}{b_i}\right)
\right]^2
\right\}^{1/2},
\end{equation}
where
\begin{equation}
F_{\ell}^i = \frac{9}{\ell^2} \frac{D_i^2(D_s-D_i)^2}{D_s^2}
\left(\frac{d z}{d D} n_{i} \right)^{-2} \frac{\Omega_m^2 H_0^4}{a_i^2}
C_{\ell}^{i \; gg}.
\end{equation}
An estimation of errors using the redshift distortion measurements
of bias can be found in Fig.~5 of Ref.~\cite{Song:2008xd}.

In the second test, it is required to measure the power spectrum
of the velocity divergence. One of the most promising ways is
to use the redshift distortions. Even though the clustering of
galaxies in real space to have no preferred direction, galaxy
maps in redshift space show an anisotropic distribution due to
the peculiar velocities of galaxies. Thus the measurements of
anisotropies allow us to extract the peculiar velocities. The
errors on the reconstruction of $P_{\theta_m \theta_m}$ from
redshift distortions are estimated
in Ref.~\cite{White:2008jy}. Although it is challenging to measure
$P_{\theta_m \theta_m}$ from the redshift surveys, it is
shown that future surveys such as Euclid can give the fractional
errors on $P_{\theta_m \theta_m}$ with a few $\%$ to 10$\%$ accuracy
depending on scales for $k < 0.1$ Mpc$^{-1}$. This is one of the major systematic errors
in the second test and we should probably wait for the surveys
like SKA to achieve sub $\%$ accuracy.

\section{Conclusion}
In this paper, we proposed a consistency test of general relativity by
combining geometrical tests and structure formation tests using
large-scale structure. From the
geometrical test using supernovae, cosmic microwave background
and baryon acoustic oscillations, we can reconstruct the Hubble
parameter $H$. From structure formation, we measure the density perturbation
$\delta_t$, the peculiar velocity function $V_m$ and the lensing potential $\Phi_-$.
We constructed a consistency test which is written only in terms of these
observables, Eq.~(\ref{eq:GRcon}). The major advantage of this approach is that
we do not need to assume any theory for modified gravity models to test
GR on cosmological scales. This test is essentially the same as the
consistency relation derived in \cite{Jain:2007yk}. We have shown that this test
probes the difference between the Newton constants in the background cosmology
and in large-scale structure. The main obstacle in realizing this test
was that we should measure all the quantities in the
consistency test at the same time and in the same location. This is not straightforward
because weak lensing measures an integrated effect of the lensing potential at
different redshifts and it is impossible to reconstruct $\Phi_-$ at a
given redshift. In this paper, we proposed a way to overcome this
problem by reconstructing the lensing potential from measured
density perturbations and peculiar velocities. In the reconstruction,
we use the Poisson equation in GR to reconstruct $\Phi$ from the density
perturbation. Thus any inconsistency between the reconstructed
lensing power spectrum and the measured lensing spectrum indicates the
failure of the Poisson equation in GR.

In summary, we have shown that, given the
assumptions (i) the cosmological principle,
(ii) energy-momentum conservation for dark matter and baryon,
(iii) existence of a Newtonian limit in the MG model, we can test general
relativity only by using the measured quantities from the background
expansion history and large-scale structure
of the Universe. We should also emphasize that there exist independent ways
to test these assumptions from observations.

We summarized the major systematic errors in the consistency
test in section V. Some of them have been studied in detail in
Ref.~\cite{Song:2008xd}, but clearly controlling these systematic
errors are essential to perform the consistency test.

\begin{acknowledgments}
We would like to thank David Bacon, Ben Hoyle, Roy Maartens, Robert Nichol
and Will Percival for helpful discussions. We are grateful to Roy Maartens
for careful reading of the manuscript. This work is supported by STFC.
\end{acknowledgments}

%\bibliography{MGVDE-V2}

\end{document}